\documentstyle[sprocl]{article}
\input{psfig}
\def\NPB{{\em Nucl. Phys.} B}
\def\PLB{{\em Phys. Lett.}  B}
\def\PRL{{\em Phys. Rev. Lett.}~}
\def\PRD{{\em Phys. Rev.} D}
\def\ZPC{{\em Z. Phys.} C}

\def\vep{\varepsilon}
\def\bea{\begin{eqnarray}}
\def\eea{\end{eqnarray}}

\def\be{\begin{eqnarray}}
\def\en{\end{eqnarray}}
\def\ra{\rangle}
\def\la{\langle}
\def\non{\nonumber}
\def\bi{\bibitem}
\def\nc{N_c^{\rm eff}}
\def\ov{\overline}
\def\lsim{ {\ \lower-1.2pt\vbox{\hbox{\rlap{$<$}\lower5pt\vbox{\hbox{$\sim$}
}}}\ } }
\def\gsim{ {\ \lower-1.2pt\vbox{\hbox{\rlap{$>$}\lower5pt\vbox{\hbox{$\sim$}
}}}\ } }
\def\vma{{_{V-A}}}

\begin{document}
\title{HADRONIC CHARMLESS $B$ DECAYS AND NONFACTORIZABLE EFFECTS}
\author{ HAI-YANG CHENG }
\address{Institute of Physics, Academia Sinica, Taipei, Taiwan 115, ROC}
\maketitle\abstracts{Hadronic charmless $B$ decays and their nonfactorizable
effects are reviewed.}
\section{Introduction}
A remarkable progress in the study of exclusive charmless $B$ decays has 
been made recently. Experimentally, CLEO \cite{CLEO} has discovered many 
new two-body decay modes 
\be
 B\to\eta' K^\pm,~\eta' K^0_S,~\pi^\pm K^0_S,~\pi^\pm K^\mp,~\omega K^\pm,
\en
and a possible evidence for $B\to \phi K^*$.
Moreover, CLEO has provided new improved upper limits for many other decay
modes. Some of the CLEO data are surprising from the theoretical point
of view: The measured 
branching ratios for $B^\pm\to\eta' K^\pm$ and $B^\pm\to\omega K^\pm$ are
about several times larger than the naive theoretical estimate.

\section{Difficulties with naive factorization}
The relevant effective weak Hamiltonian for hadronic weak $B$ decay is of
the form
\be
{\cal H}_{\rm eff}={G_F\over\sqrt{2}}\left[ \lambda_u(c_1O_1^u+c_2O_2^u)
+\lambda_c(c_1O_1^c+c_2O_2^c)-\lambda_t\sum_{i=3}^{10} c_iO_i\right],
\en
where $\lambda_i=V_{ib}V^*_{iq}$, $O_{3-6}$ are the QCD penguin operators and
$O_{7-10}$ the electroweak penguin operators. 
The Wilson coefficients $c_i(\mu)$ in Eq.~(2) have been evaluated 
to the next-to-leading order (NLO) and they depend on the choice of the
renormalization scheme. The mesonic matrix
elements are customarily evaluated using the factorization hypothesis.
Under this assumption, the 3-body hadronic matrix element $\la M_1M_2|
O|B\ra$ is approximated as the product of two matrix elements
$\la M_1|J_{1\mu}|0\ra$ and $\la M_2|J_2^\mu|B\ra$.
Although this approach for matrix elements is very simple, it encounters two
major difficulties. First, the hadronic matrix element under factorization is
renormalization scale $\mu$ independent as the vector or axial-vector
current is partially conserved. Consequently, the amplitude $c_i(\mu)
\la O\ra_{\rm fact}$ is not truly physical as the scale dependence of
Wilson coefficients does not get compensated from the matrix elements.
Second, in the
naive factorization approach, the relevant Wilson coefficient functions
for color-allowed external $W$-emission (or so-called ``class-I") and 
color-suppressed (class-II) internal $W$-emission
amplitudes are given by $a_1=c_1+c_2/N_c$, $a_2=c_2+c_1/N_c$, respectively, 
with $N_c$ the number of colors. However, naive factorization fails to 
describe class-II
decay modes. For example, the ratio $R=\Gamma(D^0\to\overline K^0\pi^0)/
\Gamma(D^0\to K^-\pi^+)$ is predicted to be $\sim {1\over 50}$, while 
experimentally \cite{PDG} $R=0.51\pm 0.07$. This implies that it is
necessary to take into account nonfactorizable contributions to the 
decay amplitude in order to render
the color suppression of internal $W$-emission ineffective.
\subsection{Scale and scheme independence of physical amplitudes}
Under the factorization hypothesis, we would like to know if it is
possible to obtain physical amplitudes independent of the choice of
the renormalization scale and scheme. The answer is yes. 
The scale and scheme dependence of the hadronic matrix elements can
be calculated in perturbation theory at the one-loop level 
\cite{Buras92,Kramer,Ali}. Schematically,
\be
\la O(\mu)\ra=g(\mu)\la O\ra_{\rm tree}, \qquad
\la {\cal H}_{\rm eff}\ra
=c^{\rm eff}\la O\ra_{\rm tree},
\en
with $g(\mu)$ being the perturbative corrections to the four-quark operators
renormalized at the scale $\mu$. Formally, one can show that 
$c^{\rm eff}=g(\mu)c(\mu)$ is $\mu$ and renormalization scheme independent. It 
is at this stage that the factorization approximation is applied to 
the hadronic matrix elements of the operator $O$ at tree level. The
physical amplitude obtained in this manner is guaranteed to be renormalization
scaheme and scale independent.
\footnote{This formulation is different from the one advocated in 
\cite{Neubert} in which the $\mu$ dependence of the Wilson coefficients
$c_i(\mu)$ are canceled out by that of the nonfactorization parameters
$\vep_8(\mu)$ and $\vep_1(\mu)$ so that the effective parameters $a_i^{\rm
eff}$ are $\mu$ independent.}

   The penguin-type corrections to $g(\mu)$ depend on $k^2$, the
gluon's momentum squared, so are the effective Wilson coefficient functions. 
To NLO, we obtain \cite{CT98}
\be
&& {c}_1^{\rm eff}=1.149, \qquad\qquad\qquad\qquad\qquad {c}_2^{\rm eff}=
-0.325,   \non \\
&& {c}_3^{\rm eff}=0.0211+i0.0045, \qquad\qquad\quad\,{c}_4^{\rm eff}=-0.0450
-i0.0136, \non  \\
&& {c}_5^{\rm eff}=0.0134+i0.0045, \qquad\qquad\quad\,{c}_6^{\rm eff}=-0.0560
-i0.0136, \non  \\
&& {c}_7^{\rm eff}=-(0.0276+i0.0369)\alpha, \qquad \quad {c}_8^{\rm eff}=
0.054\,\alpha, \non  \\
&& {c}_9^{\rm eff}=-(1.318+i0.0369)\alpha, \qquad \quad ~\,{c}_{10}^{\rm eff}=
0.263\,\alpha,
\en
at $k^2=m^2_b/2$. 

\subsection{Generalized factorization}
Because there is only one single form factor (or Lorentz scalar) 
involved in the class-I or class II decay amplitude of $B\to PP,~PV$ decays, 
the effects of nonfactorization can be lumped into the
effective parameters $a_1$ and $a_2$ \cite{Cheng94}:
\be
a_1^{\rm eff}=c_1^{\rm eff}+c_2^{\rm eff}\left({1\over N_c}+\chi_1\right),
\qquad a_2^{\rm eff}=c_2^{\rm eff}+c_1^{\rm eff}\left({1\over N_c}
+\chi_2\right),
\en
where $\chi_i$ are nonfactorizable terms and receive main contributions from
the color-octet current operators. 
Since $|c_1/c_2|\gg 1$, it is evident from 
Eq.~(5) that even a small amount of nonfactorizable contributions will have a
significant effect on the color-suppressed class-II amplitude.
If $\chi_{1,2}$ are universal (i.e. process independent) in
charm or bottom decays, then we still have a generalized factorization scheme
in which the decay amplitude is expressed in terms of factorizable 
contributions multiplied by the universal effective parameters
$a_{1,2}^{\rm eff}$. For $B\to VV$ decays, this new factorization implies
that nonfactorizable terms contribute in equal weight to all partial wave
amplitudes so that $a_{1,2}^{\rm eff}$ {\it can} be defined.
It should be stressed that, contrary to the
naive one, the improved factorization does incorporate nonfactorizable
effects in a process independent form.
Phenomenological analyses
of two-body decay data of $D$ and $B$ mesons indicate that while
the generalized factorization hypothesis in general works reasonably well, 
the effective parameters $a_{1,2}^{\rm eff}$ do show some variation from
channel to channel, especially for the weak decays of charmed mesons
 \cite{Cheng94,Kamal96,Cheng96}.
An eminent feature emerged from the data analysis is that $a_2^{\rm eff}$ 
is negative in charm decay, whereas it becomes positive in the two-body decays
of the $B$ meson \cite{Cheng94,CT95,Neubert}:
\be
a_2^{\rm eff}(D\to\overline{K}\pi)\sim -0.50\,, \qquad a_2^{\rm eff}(B\to 
D\pi)\sim 0.20-0.28\,.
\en
It follows that
\be
\chi_2(D\to \overline{K}\pi)\sim -0.36\,, \qquad \chi_2(B\to D\pi)\sim 
0.12-0.19\,.
\en
The observation $|\chi_2(B)|\ll|\chi_2(D)|$ is consistent with the 
intuitive picture that nonperturbative soft gluon effects become stronger when the
final-state particles move slower, allowing more time for significant
final-state interactions after hadronization \cite{Cheng94}.
Phenomenologically, it is often to treat the number of colors $N_c$ as
a free parameter to model the nonfactorizable contribution to hadronic
matrix elements and its value can be extracted from the data of two-body
nonleptonic decays. Theoretically, this amounts to
defining an effective number of colors $\nc$, called $1/\xi$ in \cite{BSW87},
by $1/N_c^{\rm eff}\equiv (1/N_c)+\chi$. It is clear from (7) that
\be
N_c^{\rm eff}(D\to \overline{K}\pi) \gg 3,\qquad N_c^{\rm eff}(B\to D\pi)
=1.8-2.2\approx 2\,.
\en

\section{Nonfactorizable effects in charmless $B$ decays}
We next study the nonfactorizabel effects in charmless rare $B$ decays.
We note that the effective Wilson coefficients appear in the factorizable 
decay amplitudes in the combinations $a_{2i}=
{c}_{2i}^{\rm eff}+{1\over N_c}{c}_{2i-1}^{\rm eff}$ and $a_{2i-1}=
{c}_{2i-1}^{\rm eff}+{1\over N_c}{c}^{\rm eff}_{2i}$ $(i=1,\cdots,5)$. 
As discussed in Sec.~2.2, 
nonfactorizable effects in the 
decay amplitudes of $B\to PP,~VP$ can be absorbed into the parameters
$a_i^{\rm eff}$. This amounts to replacing $N_c$ in $a^{\rm eff}_i$
by $(N_c^{\rm eff})_i$. Explicitly,
\be
a_{2i}^{\rm eff}={c}_{2i}^{\rm eff}+{1\over (N_c^{\rm eff})_{2i}}{c}_{2i-1}^{
\rm eff}, \qquad \quad a_{2i-1}^{\rm eff}=
{c}_{2i-1}^{\rm eff}+{1\over (N_c^{\rm eff})_{2i-1}}{c}^{\rm eff}_{2i}.
\en
It is customary to assume in the literature that $(N_c^{\rm eff})_1
\approx (N_c^{\rm eff})_2\cdots\approx (N_c^{\rm eff})_{10}$;
that is, the nonfactorizable
term is usually assumed to behavor in the same way in penguin and non-penguin
decay amplitudes. A closer investigation shows 
that this is not the case. We have argued in \cite{CT98} that
nonfactorizable effects in the matrix
elements of $(V-A)(V+A)$ operators are {\it a priori} different from that of 
$(V-A)(V-A)$ operators. One reason is that
the Fierz transformation of the $(V-A)(V+A)$ operators $O_{5,6,7,8}$
is quite different from that of $(V-A)(V-A)$ operators $O_{1,2,3,4}$
and $O_{9,10}$. As a result, contrary to the common assumption,
$\nc(LR)$ induced by the $(V-A)(V+A)$ operators are
theoretically different from $\nc(LL)$ generated by
the $(V-A)(V-A)$ operators \cite{CT98}.
Hence, it is plausible to assume that
\be
&& N_c^{\rm eff}(LL)\equiv
\left(N_c^{\rm eff}\right)_1\approx\left(N_c^{\rm eff}\right)_2\approx
\left(N_c^{\rm eff}\right)_3\approx\left(N_c^{\rm eff}\right)_4\approx
\left(N_c^{\rm eff}\right)_9\approx
\left(N_c^{\rm eff}\right)_{10},   \non\\
&& N_c^{\rm eff}(LR)\equiv
\left(N_c^{\rm eff}\right)_5\approx\left(N_c^{\rm eff}\right)_6\approx
\left(N_c^{\rm eff}\right)_7\approx
\left(N_c^{\rm eff}\right)_8,
\en
and that $N_c^{\rm eff}(LR)\neq N_c^{\rm eff}(LL)$. In principle, 
$N_c^{\rm eff}$ can vary from channel to channel, as in the case of charm
decay. However, in the energetic two-body $B$ decays, $\nc$
is expected to be process insensitive as supported by data 
\cite{Neubert}. 
\subsection{Classification of charmless $B$ decays}
  By studying the $\nc$-dependence of the effective parameters 
$a_i$'s (for simplicity, we will drop the superscript ``eff" henceforth),
we learn that (i) the dominant coefficients are 
$a_1,\,a_2$ for
current-current amplitudes, $a_4$ and $a_6$ for QCD penguin-induced
amplitudes, and $a_9$ for electroweak penguin-induced amplitudes,
and (ii) $a_1,a_4,a_6$ and $a_9$ are $\nc$-stable, while the other
coefficients depend
strongly on $\nc$. Therefore, for those charmless $B$ decays whose decay
amplitudes depend dominantly on $\nc$-stable coefficients, their decay
rates can be reliably predicted within the factorization approach even in
the absence of information of nonfactorizable effects.
By contrast, the decay modes involving the coefficients $a_2,a_3$ and $a_5$
are sensitive to $\nc$ and hence the nonfactorizable effects.

  In order to study hadronic charmless $B$ decays, it is useful to classify
the decay modes into several different categories. Besides the widely used 
three classes I, II, III for tree-dominated decay modes,
the penguin-dominated charmless rare $B$ decays also can be classified 
into three classes:
\begin{itemize}
\item Class-IV for those decays whose amplitudes are governed by the
parameters $a_4$ and $a_6$ in the combination $a_4+Ra_6$, where the
coefficient $R$ arises from the $(S-P)(S+P)$ part of the operator $O_6$.
In general, $R=2m_{P_b}^2/[(m_1+m_2)(m_b-m_3)]$ for $B\to P_aP_b$ with the
meson $P_b$ being factored out in the factorizable approximation,
$R=-2m_{P_b}^2/[(m_1+m_2)(m_b+m_3)]$ for $B\to V_aP_b$, and $R=0$ for
$B\to P_aV_b$ and $B\to V_aV_b$ with $V_b$ being factorizable.
Note that $a_4$ is always accompanied
by $a_{10}$, and $a_6$ by $a_8$. Examples are $\ov B_d\to K^-\pi^+,\,\ov K^0
\pi^0,\,B^-\to K^-\pi^0,\,\ov B_s\to K^+K^-,\,K^0
\ov K^0,\cdots$.
\item Class-V modes for those decays whose amplitudes are governed by the
effective coefficients $a_3,a_5,a_7$ and
$a_9$ in the combinations $a_3\pm a_5$ and/or $a_7\pm a_9$.  
Examples are $\ov B_d\to \phi\pi^0,\,B^-\to\phi\pi^-,\,\ov B_s\to\phi\pi^0$.
\item Class-VI involving the interference of class-IV and class-V decays,
e.g. $B\to K\eta',\,K\omega,\,K\phi$  ($B=B_u,B_d,B_s$).
\end{itemize}

    Sometimes the tree and penguin contributions are comparable.
For example, decays $\ov B_s\to K^0\omega,K^{*0}\omega$ fall into 
the classes of II and VI.

\subsection{Some general features for penguin-dominated proceses}
For penguin-dominated decay modes, some observations can be made:
\begin{itemize}
\item For class-IV modes, the decay rates obey the pattern:
\be
\Gamma(B\to P_aP_b)>\Gamma(B\to P_aV_b)\sim\Gamma(B\to V_a V_b)>\Gamma
(B\to V_aP_b),
\en
where $M=P_b$ or $V_b$ is factorizable under the factorization assumption.
For example, the branching ratios for $\ov B^0\to K^-\pi^+,K^{*-}\pi^+,K^{*-}
\rho^+,K^-\rho^+$ are predicted to be $\sim 1.5\times 10^{-5},\,0.7\times 
10^{-5},\,0.6\times 10^{-5}$ and $0.5\times 10^{-6}$ respectively.
This hierarchy follows from various
interference between the penguin terms characterized by the effective
coefficients $a_4$ and $a_6$. On the contrary, in general
$\Gamma(B\to P_aV_b)>\Gamma(B\to P_aP_b)$ for tree-dominated decays
because the vector meson has three different polarization states.
\item Among the two-body charmless $B$ decays,
the class-III decay modes $B^-\to \eta' K^-,\,\ov B_d\to \eta' K^0$ and $\ov
B_s\to \eta\eta',\eta'\eta'$ have the largest branching ratios. Theoretically,
${\cal B}(B^-\to\eta' K^-)\approx {\cal B}(\ov B_d\to\eta' K^0)\sim
4\times 10^{-5}$ and ${\cal B}(\ov B_s\to\eta\eta',\eta'\eta')\sim 2\times 
10^{-5}$. These decay modes receive two different sets of penguin terms
proportional to $a_4+Ra_6$ with $R>0$. By contrast, $VP,\,VV$ modes in
charm decays or bottm decays involving charmed mesons usually have 
larger branching ratios than the $PP$ mode.
\item The decay amplitudes of $B\to M\pi^0,M\rho^0$ with $\pi^0(\rho^0)$
being factored out contain the electroweak penguin contributions proportional 
to ${3\over 2}(-a_7+a_9)X_u^{(BM,\pi^0)}$ and ${3\over 2}(a_7+a_9)
X_u^{(BM,\rho^0)}$, respectively, \cite{CCT} with
\be
X_u^{(BM,\pi^0)}=\la \pi^0|(\bar uu)_\vma|0\ra\la M|(\bar qb)_\vma|\ov
B\ra.
\en
For $\ov B_s\to\eta'\pi,\eta'\rho,\phi\pi,\phi\rho$, QCD penguin 
contributions are canceled out so that these decays are dominated by 
electroweak penguins. Hence, a measurement of them can be used to determine
the effective coefficient $a_9$.
\footnote{It has been suggetsed in \cite{Ali98} that $\ov B_d\to\ov 
K^0\rho^0$ can be utilized to extract $a_9$. However, this method 
relies on the cancellation between the QCD penguin terms characterized by
$a_4$ and $a_6$. In general, this
cancellation is not complete and this makes this decay mode less clean
than $\ov B_s\to(\eta',\phi)(\pi,\rho)$ for determining $a_9$.}

\end{itemize}

\subsection{Nonfactorizable effects in spectator amplitudes}
We focus on class-III decay modes dominated by the spectator diagrams induced
by the current-current operators $O_1$ and $O_2$ and are sensitive to 
the interference between external and internal $W$-emission amplitudes. 
Good examples are the class-III modes: $B^\pm\to \omega\pi^\pm,~\pi^0
\pi^\pm,~\eta\pi^\pm,~\pi^0\rho^\pm,\cdots$, etc. Considering $B^\pm\to
\omega\pi^\pm$, we find that
the branching ratio is sensitive to $1/\nc$ and has the 
lowest value of order $2\times 10^{-6}$ at $\nc=\infty$ and then increases 
with $1/\nc$. The 1997 CLEO measurement yields \cite{CLEOomega1}
\be
{\cal B}(B^\pm\to\omega\pi^\pm)=\left(1.1^{+0.6}_{-0.5}\pm 0.2\right)
\times 10^{-5}.
\en
Consequently, $1/\nc>0.35$ is preferred by the data \cite{CT98}. 
Because this decay 
is dominated by tree amplitudes, this in turn implies that $\nc(V-A)<2.9$.
If the value of $\nc(V-A)$ is fixed to be 2, the branching ratio 
for positive $\rho$, which is preferred by the current analysis 
\cite{Parodi}, will be of order $(0.9-1.0)\times 10^{-5}$, which is 
very close to the central value of the measured one. Unfortunately,
the significance of $B^\pm\to\omega\pi^\pm$ is reduced in the
recent CLEO analysis and only an upper limit is quoted \cite{CLEOomega2}:
${\cal B}(B^\pm\to\pi^\pm\omega)<2.3\times 10^{-5}$.
Since ${\cal B}(B^\pm\to K^\pm\omega)=(1.5^{+0.7}_{-0.6}\pm 0.2)
\times 10^{-5}$ and ${\cal B}(B^\pm\to h^\pm\omega)=(2.5^{+0.8}_{-0.7}\pm 0.3)
\times 10^{-5}$ with $h=\pi,~K$, the central value of ${\cal B}(B^\pm\to
\pi^\pm\omega)$ remains about the same as (11). The fact that $\nc(LL)\sim 2$ 
is preferred in charmless two-body decays of the $B$ meson 
is consistent with the nonfactorizable term extracted from $B\to (D,D^*) 
\pi,~D\rho$ decays, namely $\nc(B\to D\pi)\approx 2$. 
Since the
energy release in the energetic two-body decays $B\to\omega\pi$, $B\to D\pi$
is of the same order of magnitude, it is thus expected that $\nc(LL)|_{B
\to\omega\pi}\approx 2$.

Just like the decay $B^-\to\pi^-\omega$,
the branching ratio of $B^-\to\pi^-\pi^0$ also increases with $1/\nc$.
The CLEO measurement is \cite{CLEOpik}
\be
{\cal B}(B^\pm\to\pi^\pm\pi^0)=\left(0.9^{+0.6}_{-0.5}\right)\times
10^{-5}~~<2.0\times 10^{-5}.
\en
However, the errors are so large that it is meaningless to put a 
sensible constraint 
on $\nc(LL)$. Nevertheless, we see that in the range \cite{Ali} 
$0\leq 1/\nc\leq 0.5$, $\nc(LL)\approx 2$ is most favored.

 In analogue to the decays $B\to D^{(*)}\pi(\rho)$, the interference effect of
spectator amplitudes in class-III charmless $B$ decay can be tested 
by measuring the ratios:
\be  \label{ratios}
R_1\equiv 2\,{{\cal B}(B^-\to\pi^-\pi^0)\over {\cal B}(\bar B^0\to \pi^-\pi^+
)},~~ R_2\equiv 2\,{{\cal B}(B^-\to\rho^-\pi^0)\over {\cal B}(\bar 
B^0\to \rho^-\pi^+)},~~ R_3\equiv 2\,{{\cal B}(B^-\to\pi^-\rho^0)\over 
{\cal B}(\bar B^0\to \pi^-\rho^+)}.
\en
Evidently, the ratios $R_i$ are greater (less) than unity when the 
interference is constructive (destructive). Numerically we find
\be
R_1=\cases{1.74,  \cr 0.58, \cr} \qquad R_2=\cases{1.40, \cr 0.80, \cr} \qquad
R_3=\cases{2.50 & for~$\nc=2$,  \cr  0.26 & for~$\nc=\infty$. \cr}
\en
Hence, a measurement of
$R_i$ (in particular $R_3$), which has the advantage of being independent of
the Wolfenstein parameters $\rho$ and $\eta$, will constitute a very 
useful test on the
effective number of colors $\nc(LL)$. The present experimental 
information on $\overline{B}^0\to\pi^+\pi^-$ is \cite{CLEOpik}
\be
{\cal B}(\overline{B}^0\to\pi^\pm\pi^\mp)=(0.7\pm 0.4)\times 10^{-5}~~<1.5
\times 10^{-5}.
\en
As far as the experimental central value of $R_1$ is concerned, it appears 
that $1/\nc\sim
0.5$ is more favored than any other small values of $1/\nc$.

\subsection{Nonfactorizable effects in penguin amplitudes}
The penguin amplitude of the class-VI mode $B\to \phi K$ is proportional to
$(a_3+a_4+a_5)$ and hence sensitive to the variation of $\nc$. 
Neglecting $W$-annihilation and space-like penguin diagrams, 
we find \cite{CT98} that $\nc(LR)=2$
is evidently excluded from the present CLEO upper limit \cite{CLEOomega2}
\be \label{phiK}
{\cal B}(B^\pm\to\phi K^\pm)< 0.5\times 10^{-5},
\en
and that $1/\nc(LR)<0.23$ or $\nc(LR)>4.3\,$.
A similar observation was also made in \cite{Deshpande2}. 
The branching ratio of $B\to\phi K^*$, the average of $\phi K^{*-}$ and
$\phi K^{*0}$ modes, is also measured recently by CLEO with the result
\cite{CLEOomega2}
\be    \label{phiK*}
{\cal B}(B\to\phi K^*) 
=\left(1.1^{+0.6}_{-0.5}\pm 0.2\right)\times 10^{-5}.
\en
We find that the allowed region for $\nc(LR)$ is $4\gsim \nc(LR)\gsim 1.4$.
This is in contradiction to the constraint
$\nc(LR)>4.3$ derived from $B^\pm\to\phi K^\pm$. In fact,  
the factorization approach predicts that
$\Gamma(B\to\phi K^*)\approx\Gamma(B\to\phi K)$ 
when the $W$-annihilation type of contributions is neglected. The current
CLEO measurements (\ref{phiK}) and (\ref{phiK*}) are obviously not consistent 
with the
prediction based on factorization. One possibility is that
generalized factorization is not applicable to $B\to VV$.
Therefore, the discrepancy between ${\cal B}(B\to\phi K)$ and 
${\cal B}(B\to\phi K^*)$ will measure
the degree of deviation from the generalized factorization that has been
applied to $B\to\phi K^*$. It is also possible that the absence 
of $B\to\phi K$ events is a downward fluctuation of the
experimental signal. At any rate, 
in order to clarify this issue and to pin down the effective number of 
colors $\nc(LR)$, we urgently need 
measurements of $B\to\phi K$ and $B\to\phi K^*$, especially the neutral 
modes, with sufficient accuracy. 

   The decay mode $B\to\eta' K$ also provides another useful information on 
$\nc(LR)$. The discrepancy between the experimental measurements
\be
{\cal B}(B^\pm\to\eta' K^\pm) &=& \left(6.5^{+1.5}_{-1.4}\pm 0.9\right)\times
10^{-5}, \non \\
 {\cal B}(B^0\to\eta' K^0) &=& \left(4.7^{+2.7}_{-2.0}\pm 0.9
\right)\times 10^{-5}
\en
and the theoretical estimates \cite{Chau1,Kramer,Du} of order $1\times 
10^{-5}$ 
seems to call for some new mechanisms unique to the $\eta'$ production or 
even some new physics beyond the Standard Model.

\begin{figure}[ht]
\vspace{2cm}
    \centerline{\psfig{figure=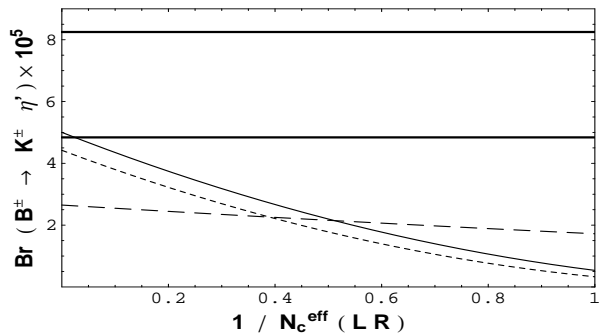,width=7.5cm,height=4.0cm}}
\vspace{-1.2cm}
    \caption{{\small The branching ratio of $B^\pm\to\eta' K^\pm$ as a
   function of $1/\nc(LR)$ with $\nc(LL)$ being fixed at the value of 2
   and $\eta=0.34$ and $\rho=0.16$.
   The charm content of the $\eta'$ with $f_{\eta'}^c=-6\,{\rm MeV}$ 
   contributes to the solid curve, but not to the dotted curve.
   The anomaly contribution to $\la\eta'|\bar s\gamma_5s|0\ra$
   is included. For comparison, the prediction for the case that
   $\nc(LL)=\nc(LR)$ as depicted by the dashed curve is also shown.
    The solid thick lines
   are the CLEO measurements with one sigma errors.}}
\end{figure}

   In the conventional way of treating $\nc(LR)$ and $\nc (LL)$ in the same
manner, the branching ratio of $B^\pm \to\eta' K^\pm$ can be enhanced
to the order of $(2-3)\times 10^{-5}$ due to
the small running strange quark mass at $\mu=m_b$ and $SU(3)$ breaking in
the decay constants $f_8$ and $f_0$ (corresponding to the dashed curve 
in Fig.~1). It should be emphasized that this
prediction has taken into account the anomaly effect in the matrix 
element $\la\eta' |\bar s\gamma_5s|0\ra$. Specifically, 
\be
\la\eta'|\bar s\gamma_5 s|0\ra=-i{m_{\eta'}^2\over 2m_s}\,\left(f_{\eta'}^s
-f^u_{\eta'}\right),
\en
where the QCD anomaly effect is manifested by the decay constant $f_{\eta'}
^u$. Since $f_{\eta'}^u\sim {1\over 2}f_{\eta'}^s$ and the decay
amplitude is dominated by $(S-P)(S+P)$ matrix elements, it is obvious that
the decay rate of $B\to\eta' K$ would be (wrongly) enhanced considerably 
in the absence of the anomaly term in $\la\eta'|\bar s\gamma_5 s|0\ra$.

   It has been advocated that the new internal $W$-emission contribution 
coming from the Cabibbo-allowed process $b\to c\bar c s$ followed 
by a conversion of the $c\bar c$ pair into the $\eta'$ via two gluon 
exchanges may play an important role
since its mixing angle $V_{cb}V_{cs}^*$ is as large as that
of the penguin amplitude and yet its Wilson coefficient $a_2$ 
is larger than that of penguin operators. 
The decay constant $f_{\eta'}^c$, 
defined by $\la 0|\bar c\gamma_\mu\gamma_5c|\eta'\ra=if_{\eta'}^c
q_\mu$, has been estimated to be $f_{\eta'}^c=(50-180)$ MeV,
based on the OPE, large-$N_c$ approach and QCD low energy 
theorems \cite{Halperin}. Recent refined estimates \cite{Ali2,Petrov} 
give $f_{\eta'}^c=-(2\sim15)$ MeV, which is in strong contradiction 
in magnitude and sign to the estimate of \cite{Halperin}.
It turns out that if $\nc(LL)$ is treated to be the same as $\nc(LR)$,
this new mechanism is not welcome for explaining
${\cal B}(B\to\eta' K)$ at small $1/\nc$ due to the fact that its 
contribution is proportional to $a_2$, which is negative at small $1/\nc$.

   We have shown in \cite{CT98} that if $\nc(LL)\sim 2$ and 
$\nc(LR)>\nc(LL)$, ${\cal B}(B^\pm\to\eta' K^\pm)$ at $1/\nc(LR)\leq 0.2$ 
will be enhanced considerably from $(2.5-3)\times 10^{-5}$ to $(3.7-5)
\times 10^{-5}$ (see Fig.~1). First, the $\eta'$ charm content contribution
now contributes in the right direction to the decay rate 
irrespective of the value of $\nc(LR)$ as $a_2$ now is always positive. 
Second, the interference between
the spectator amplitudes of $B^\pm\to\eta' K^\pm$ is constructive. 
Third, the term proportional to
$2(a_3-a_5)X_u^{(BK,\eta')}+(a_3+a_4-a_5)X_s^{(BK,\eta')}$
is enhanced when $(\nc)_3=(\nc)_4=2$. 
The agreement with experiment for $B^\pm\to\eta' K^\pm$ thus provides another 
strong support for $\nc(LL)\sim 2$ and for
the relation $\nc(LR)>\nc(LL)$.

\section{Final-state interactions and $B\to\omega K$}
  The CLEO observation \cite{CLEOomega2} of a large branching ratio for 
$B^\pm\to \omega K^\pm$
\be
{\cal B}(B^\pm\to\omega K^\pm)=\left(1.5^{+0.7}_{-0.6}\pm 0.2\right)
\times 10^{-5},
\en
is difficult to explain at first sight. Its factorizable amplitude is
of the form
\be  \label{omegaK}
A(B^-\to\omega K^-) \propto (a_4+Ra_6)X^{(B\omega, K)}+(2a_3+2a_5+{1\over 2}
a_9)X^{(BK,\omega)}+\cdots,
\en
with $R=-2m_K^2/(m_bm_s)$, where ellipses represent for contributions from 
$W$-annihilation and space-like penguin diagrams.
It is instructive to compare this decay mode closely with $B^-\to\rho K^-$ 
\be   \label{rhoK}
A(B^-\to\rho^0 K^-) \propto (a_4+Ra_6)X^{(B\rho,K)}+{3\over 2}
a_9X^{(BK,\rho)}+\cdots.
\en
Due to the destructive interference between $a_4$ and $a_6$ penguin terms,
the branching ratio of $B^\pm\to\rho^0 K^\pm$ is estimated to be of order
$5\times 10^{-7}$. The question is then why is the observed rate of the
$\omega K$ mode much larger than the $\rho K$ mode ? By comparing 
(\ref{omegaK}) with (\ref{rhoK}),
it is natural to contemplate that the penguin contribution
proportional to $(2a_3+2a_5+{1\over 2}a_9)$ accounts for the large
enhancement of $B^\pm\to \omega K^\pm$. However, this is not the case: The
coefficients $a_3$ and $a_5$, whose magnitudes are smaller than $a_4$ 
and $a_6$, are not large enough to accommodate the data unless $\nc(LR)
<1.1$ or $\nc(LR)>20$ (see Fig.~9 of \cite{CT98}).

   So far we have neglected three effects in the consideration of 
$B^\pm\to\omega K^\pm$: $W$-annihilation, space-like
penguin diagrams and final-state interactions (FSI).
It turns out that FSI may play the dominant role for $B^\pm\to\omega K^\pm$.
The weak decays $B^-\to
K^{*-}\pi^0$ via the penguin process $b\to su\bar u$ and $B^-\to 
K^{*0}\pi^-$ via $b\to sd\bar d$ followed by the quark rescattering 
reactions $\{K^{*-}
\pi^0,~K^{*0}\pi^-\}\to\omega K^-$ contribute constructively to $B^-\to
\omega K^-$, but destructively to $B^-\to\rho K^-$.
Since the branching ratios for $B^-\to K^{*-}\pi^0$ and
$K^{*0}\pi^-$ are large, of order $(0.5-0.8)\times 10^{-5}$, it is 
conceivable that a large branching ratio for $B^\pm\to\omega K^\pm$ can 
be achieved from FSI via inelastic scattering. Moreover, if FSI dominate,
it is expected that ${\cal B}(B^\pm\to\omega K^\pm)\approx (1+\sqrt{2})^2
{\cal B}(B^0\to\omega K^0)$.

\section{Conclusions}
  To summarize, the CLEO data of $B^\pm\to\omega\pi^\pm$ 
available last year clearly indicate that $\nc(LL)$ is favored to be small, 
$\nc(LL)<2.9\,$. This is consistent with the observation that 
$\nc(LL)\approx 2$ in $B\to D\pi$ decays. Unfortunately, the significance 
of $B^\pm\to\omega\pi^\pm$ is reduced in the recent CLEO analysis and only an
upper limit is quoted. Therefore, a measurement of its branching ratio is
urgently needed. In analogue to the class-III $B\to D\pi$ decays, the
interference effect of spectator amplitudes in charged $B$ decays 
$B^-\to\pi^-\pi^0,~\rho^-\pi^0,~\pi^-\rho^0$ is sensitive to $\nc(LL)$;
measurements of them [see (\ref{ratios})] will be very useful to pin down 
the value of $\nc(LL)$. 

As for $\nc(LR)$, we found that the constraints on $\nc(LR)$
derived from $B^\pm\to\phi K^\pm$ and $B\to\phi K^*$ are no consistent. 
Under the factorization hypothesis, the decays $B\to\phi K$ and 
$B\to\phi K^*$ should have almost the same branching ratios, a prediction not
borne out by current data. Therefore, it is crucial to measure the charged
and neutral decay modes of $B\to\phi(K,K^*)$ in order to see if
the generalized factorization approach is applicbale to $B\to\phi K^*$ decay.
Nevertheless, our analysis of $B\to\eta' K$ indicates that $\nc(LL)\approx 2$
is favored and $\nc(LR)$ is preferred to be larger. Since the energy 
release in the energetic two-body charmless $B$ decays is not less
than that in $B\to D\pi$ decays, it is thus expected that
\be
|\chi({\rm 2-body~rare~B~decay})|\lsim |\chi(B\to D\pi)|.
\en
It follows from Eq.~(7) that $\nc(LL)\approx \nc(B\to D\pi)\sim 2$ and
$\nc(LR)\sim 2-5$, depending on the sign of $\chi$. Therefore, we
conjecture that $\nc(LR)\sim 5>\nc(LL)\sim 2$.

\section*{Acknowledgments} 
I would like to thank the organizer Guey-Lin Lin for this well run and 
stimulating workshop. This work was supported in part by the National
Science Council of the Republic of China under Grant NSC87-2112-M006-018.

\end{document}